\documentclass[prl,superscriptaddress,showpacs,a4paper,twocolumn,floatfix]{revtex4}

\usepackage{amsmath} 
\usepackage{graphics} 
\usepackage{dcolumn} 
\usepackage{epsfig} 
\usepackage{color} 
\usepackage[colorlinks=true]{hyperref}

\graphicspath{{figures/}}

\begin{document}

\title{Orientational Ordering of Non-Planar Phthalocyanines on Cu(111):
  Strength and Orientation of the Electric Dipole Moment}

\author{A.~Gerlach} 
\author{T.~Hosokai} 

\affiliation{Institut f\"ur Angewandte Physik, Universit\"at T\"ubingen, 72076
  T\"ubingen, Germany}

\author{S.~Duhm}
\affiliation{Graduate School of Advanced Integration Science, Chiba
  University,  Chiba 263-8522, Japan} 

\affiliation{Institut f\"ur Physik, Humboldt-Universit\"at zu Berlin, 12489
  Berlin, Germany}

\author{S.~Kera}
\affiliation{Graduate School of Advanced Integration Science, Chiba
  University, Chiba 263-8522, Japan}

\author{O. T.~Hofmann}
\author{E.~Zojer}

\affiliation{Institut f\"ur Festk\"orperphysik, Technische Universit\"at Graz,
  8010 Graz, Austria}

\author{J.~Zegenhagen}

\affiliation{European Synchrotron Radiation Facility, 6 Rue Jules Horowitz,
  BP 220, 38043 Grenoble Cedex 9, France}

\author{F.~Schreiber}

\affiliation{Institut f\"ur Angewandte Physik, Universit\"at T\"ubingen, 72076
  T\"ubingen, Germany}

\date{\today}

\begin{abstract} 
  In order to investigate the orientational ordering of molecular dipoles and
  the associated electronic properties, we studied the adsorption of
  chlorogallium phthalocyanine molecules (GaClPc, Pc =
  C$_{32}$N$_{8}$H$_{16}\,^{-2}$) on Cu(111) using the X-ray standing wave
  technique, photoelectron spectroscopy, and quantum chemical calculations.
  We find that for sub-monolayer coverages on Cu(111) the majority of GaClPc
  molecules adsorb in a 'Cl-down' configuration by forming a covalent bond to
  the substrate.  For bilayer coverages the XSW data indicate a co-existence
  of the 'Cl-down' and 'Cl-up' configuration on the substrate.  The structural
  details established for both cases and supplementary calculations of the
  adsorbate system allow us to analyze the observed change of the work
  function.
\end{abstract}

\maketitle

\label{sec:intro}

The adsorption of organic semiconductor molecules has been in the focus of
numerous experimental and theoretical investigations -- many of them
addressing the subtle interplay of electronic and structural properties. Early
studies~\cite{ishii_am99}, which show that the energy levels of organic
semiconductor/metal interfaces can exhibit large deviations from the
Schottky-Mott relation, conveyed the significance of interface dipoles.  Until
today and despite the ubiquity of this concept in the field of organic
materials, the origin of the interface dipole often remains vague.

\begin{figure}[b]
  \centering
  \includegraphics[width=\columnwidth]{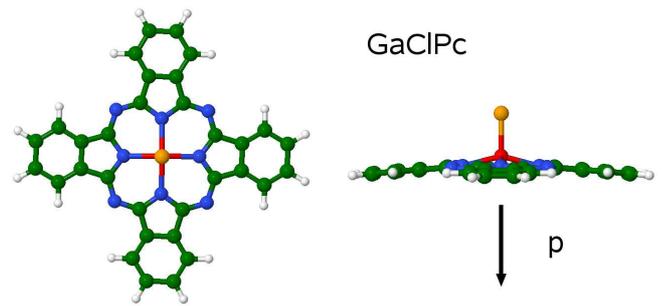}
  \caption{Structure of the free GaClPc (Pc = C$_{32}$N$_{8}$H$_{16}\,^{-2}$)
    molecule as obtained by geometry optimization using the Gaussian03 program
    package.  The results show that the Ga (red) and Cl atoms (orange) are
    located above the molecular plane with a Ga-Cl bond length of
    2.21\,\AA. The Pc group itself is non-planar with the two inequivalent N
    species (blue) 0.50\,\AA{} and 0.58\,\AA{} below the Ga atom.}
  \label{fig:gaclpc_gasphase}
\end{figure}

To establish a better understanding of the energy level alignment at the
interface one should not neglect effects related to the molecular structure of
organic adsorbates: Planar molecules such as
F$_{16}$CuPc~\cite{gerlach_prb05}, PTCDA~\cite{hauschild_prl05,gerlach_prb07}
or pentacene derivatives~\cite{koch_jacs08}, for example, can distort upon
adsorption due to the interaction with the substrate and therefore exhibit an
induced molecular dipole. Non-planar molecules such as
TiOPc~\cite{kera_prb04}, SnPc~\cite{stadler_prb06,wooley_ss07,wang_acie09},
SubPc~\cite{berner_prb03,petrauskas_jap04} and VOPc~\cite{duncan_ss10}, which
may adsorb in different orientations, form layers with at least partially
aligned dipole moments. Hence, for this class of systems the orientational
order on the surface is a quantity which strongly influences the interface
dipole. In particular, it has been shown that depending on the orientation a
layer of molecular dipoles $p$ with an area density $N_\mathrm{dip}$ can shift
the vacuum level (VL) in either direction and, therefore, increase or decrease
the work function $\Phi$ of the sample according to~\cite{fukagawa_prb06}
\begin{equation}
  \Delta \Phi_\mathrm{dip} = \pm e p N_\mathrm{dip}/\epsilon_0
  \epsilon,
  \label{eq:dipole}
\end{equation}
where $\epsilon$ is the effective dielectric constant of the monolayer.  An
experimentally and theoretically challenging model system of non-planar
organic molecules with a significant dipole moment, for which these effects
can be directly studied, is chlorogallium phthalocyanine (GaClPc,
Fig.~\ref{fig:gaclpc_gasphase})~\cite{wu_jpca04}.

In this letter, we present a detailed study on the bonding and orientational
ordering of GaClPc on Cu(111) surfaces using the X-ray standing wave (XSW)
technique~\cite{zegenhagen_ssr93}, ultra-violet photoelectron spectroscopy
(UPS) and density functional theory (DFT) based calculations.  While XSW data
are taken to determine exact the atomic positions along the surface normal
and, thereby, also the orientation of the molecules, the UPS measurements
reveal how the adsorbate affects the electronic energy levels. In particular,
our results demonstrate that the observed VL shift can be modeled very well if
one accounts for the adsorption geometry and the orientation of the molecules.

\label{sec:experiment}

The XSW experiments were carried out at beamline ID32 of the European
Synchrotron Radiation Source (ESRF) in Grenoble. Using the (111)-reflection of
the copper substrate we have taken data in back-reflection
geometry~\cite{gerlach_prb05,gerlach_prb07}, corresponding to the lattice
plane spacing $d=2.08\,$\AA{} with photon energies around
$E_\mathrm{Bragg}=2.97\,$keV.
The UPS experiments for different coverages were performed with a
photoelectron spectrometer equipped with a standard He\,I light source (photon
energy $21.2\,$eV) in our home laboratory.
The Cu(111) crystal with a small mosaicity was prepared under ultra-high
vacuum conditions by repeated cycles of argon sputtering and annealing (base
pressure $4 \times 10^{-10}\,$mbar).
The GaClPc material was purchased from Sigma-Aldrich, purified by gradient
sublimation, and thoroughly degassed. With a quartz microbalance typical
deposition rates of 0.2\,\AA/min were realized.  For all experiments reported
below the substrate temperature was carefully monitored.
To model the electronic structure of the molecules on the Cu(111) surface, we
employed slab-type DFT based band structure calculations using
VASP~\cite{kresse96_prb96}; more details on the theoretical methods and the
chosen geometry are given in the supplementary material~\cite{supp_mat}.

\label{sec:results}

\begin{figure}[htbp]
  \centering \includegraphics[width=\columnwidth]{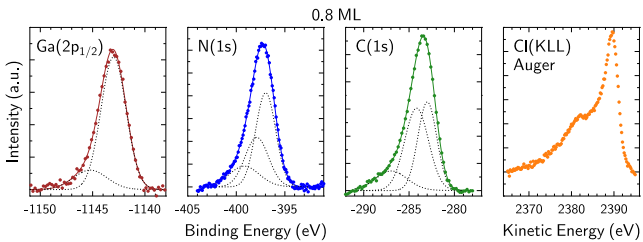}
  \caption{Background corrected core-level spectra measured with a
    hemispherical electron energy analyzer for a sub-monolayer
    of GaClPc on Cu(111). The intensity of the peaks relative to the substrate
    signal has been used to determine the surface coverage in monolayer
    equivalent units (ML).}
  \label{fig:xpsspectra}
\end{figure}

The core-level spectra of GaClPc on Cu(111) show several signals that are
suitable for XSW experiments (Fig.~\ref{fig:xpsspectra}).  By measuring
photoemission spectra related to all atomic species in the molecule, namely
C(1s), N(1s), Ga(2p$_{1/2}$), and Cl(2s)/Cl(KLL) Auger, we achieve a detailed
electronic and structural characterization of the system.

The bonding distances and orientation of the molecule have been derived from
the characteristic variation of the photoelectron yield $Y_p$ around the Bragg
condition. The XSW data, which can be analyzed according to a well-established
procedure using the experimental reflectivity $R$ of the sample, yield two
structural parameters: the coherent position $P_H$ and coherent fraction
$f_H$, both of which can be determined by a least-square fitting
routine~\cite{supp_mat}. While $P_H$ gives the average position of the atoms
relative to the diffraction planes, $f_H$ yields the spread of positions
within the ensemble.

\begin{figure*}[htbp]
  \centering 
  \includegraphics[width=\textwidth]{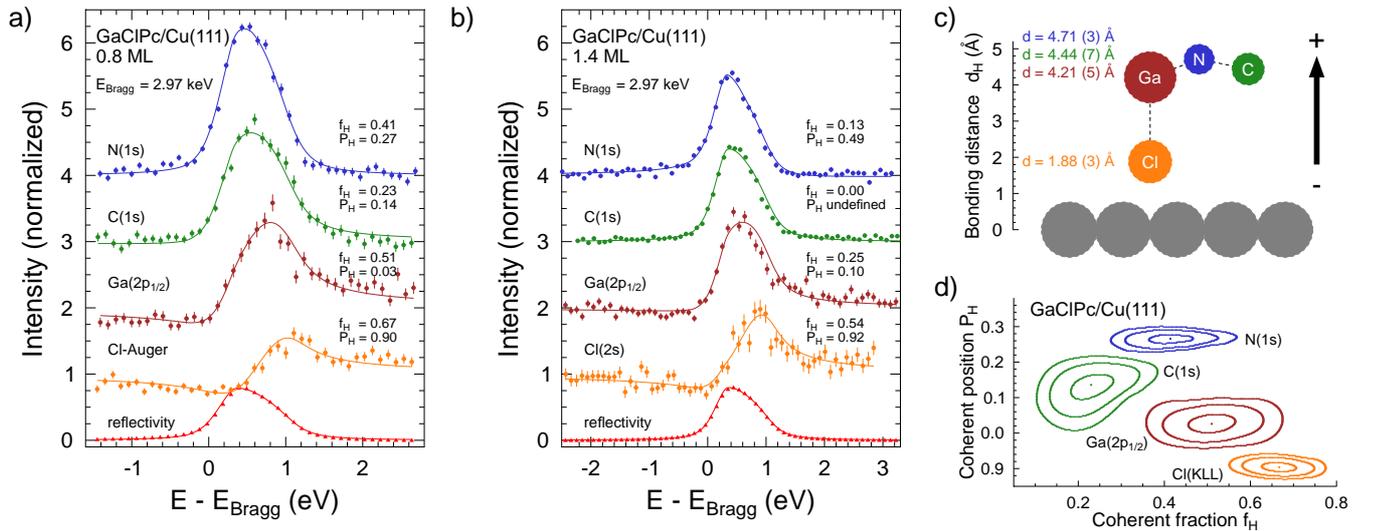}
  \caption{(a),(b) XSW data for GaClPc on Cu(111) taken for two different
    coverages in back-reflection geometry at elevated temperature. The photon
    energy has been scanned around $E_\mathrm{Bragg}=2.97\,$keV to measure the
    photoelectron yield (circles) and reflectivity (triangles). Least-square
    fits to the data (solid lines) give the coherent fraction $f_H$ and
    coherent position $P_H$ that are related to the adsorption geometry. (c)
    Average bonding distances $d_H$ for GaClPc as derived from the data in
    (a). (d) $\chi^2$-confidence map for fits to the XSW data in (a). The
    $1\sigma$-, $2\sigma$-, and $3\sigma$-contour levels show the statistical
    significance of the data analysis~\cite{gerlach_prb07}.}
  \label{fig:cu111_gaclpc1}
\end{figure*}

Figure~\ref{fig:cu111_gaclpc1} shows XSW results for two different coverages,
which were derived from series of core-level spectra.  Unlike planar organic
molecules~\cite{gerlach_prb05, gerlach_prb07}, the data corresponding to the
different atoms exhibit pronounced differences.  For sub-monolayer coverages
($0.8\,$ML, Fig.~\ref{fig:cu111_gaclpc1}a) our results indicate high
orientational order within the molecular ensemble on the surface. To convert
the coherent positions given in Fig.~\ref{fig:cu111_gaclpc1}a, i.e.\ $P_H$(Ga)
= 0.03, $P_H$(C) = 0.14, $P_H$(N) = 0.27, and $P_H$(Cl)= 0.90, into the
(average) bonding distances $d_H$ one must take into account the modulo-$d$
ambiguity of the XSW technique~\cite{zegenhagen_ssr93}. In principle, two
different configurations can be considered for low GaClPc coverages: Either a
scenario 'Cl-up' with interatomic distances which deviate strongly from the
gas phase structure (e.g.\ a hypothetical Ga-Cl bond length of only $1.83 \pm
0.07\,$\AA{} compared to $2.21\,$\AA{} in the gas phase), or a scenario
'Cl-down' with a corresponding Ga-Cl bond length of $2.33 \pm 0.07\,$\AA{} and
a vertical N-Ga distance of $0.50\,$\AA, i.e.\ values that are close to the
interatomic distances in the gas phase.
DFT calculations for the isolated molecule, which show that large molecular
distortions of GaClPc are energetically very unfavorable, can be used to rule
out the 'Cl-up' case.  The obvious conclusion that the molecules adsorb --
similar to other Pc derivatives~\cite{berner_prb03,petrauskas_jap04} --
predominantly in a 'Cl-down' configuration implies that the molecular dipoles
contribute with $\Delta \Phi_\mathrm{dip} <0$ to the overall VL shift.  Having
established the bonding distances for the low coverage
(Fig.~\ref{fig:cu111_gaclpc1}c) with the corresponding error bars
(Fig.~\ref{fig:cu111_gaclpc1}d), we see that the Cu-Cl layer spacing of
$1.88\,$\AA{} coincides with experimental results for chlorine adsorption on
Cu(111)~\cite{crapper_ss87}.  This agreement suggests a covalent bonding of
the Cl atoms to the substrate and, moreover, a hollow adsorption site as
reported in Ref.~\cite{crapper_ss87}.

Complementing this picture, we find increasing magnitudes of $f_H$ for carbon
(0.23), nitrogen (0.41), gallium (0.51), and chlorine (0.67), which indicate
surprisingly different spatial spreads.  Therefore, we studied the influence
of thermally activated librational motions of the GaClPc molecules in the
'Cl-down' configuration.  Specific model calculations based on a statistical
description of the ensemble (presented in the supplementary
material~\cite{supp_mat}) were performed to refine the analysis.  The
time-averaged spread of positions resulting from our simulations gives XSW
parameters $f_H$ and $P_H$, which -- by comparison with the experimental
values -- allow us to determine the libration amplitude.  The analysis shows
that the atomic species within the GaClPc molecule are affected very
differently and that even for small amplitudes there are strong changes of the
coherent fractions for C and N.  The relative magnitude of the coherent
fractions is consistent with a libration amplitude (rms) of $17 \pm 3^\circ$
around the essentially fixed Cl-atom.

For higher GaClPc coverages ($1.4\,$ML, Fig.~\ref{fig:cu111_gaclpc1}b) the
situation becomes more complex.  The relatively small coverage dependence
found for the Cl (and Ga) data contrasts with significant changes for the N
(and C) signal.  In particular the very different values $P_H$(N)=0.49 and
$f_H$(N)=0.13 for nitrogen cannot be explained with only one molecular
orientation.  Model calculations, which take into account that the XSW
parameter $f_H$ and $P_H$ refer to a coherent average over all inequivalent
sites in the bilayer~\cite{zegenhagen_ssr93}, i.e.\ 
\begin{equation}
  f_H e^{2 \pi i \, P_H} = \frac{1}{N} \sum_{k=1}^{N} \left( e^{2 \pi i \,
      \Delta_k^\mathrm{down} / d} + e^{2 \pi i \, \Delta_k^\mathrm{up} / d} \right) ,
  \label{eq:average}
\end{equation}
($\Delta_k^\mathrm{down/up}$ denoting the position of the $k^\mathrm{th}$ atom
relative to the substrate lattice), indicate a co-existence of 'Cl-up' and
'Cl-down' configurations on the surface. The adsorption of GaClPc molecules
with opposite orientation in the second layer -- a phenomenon reported also
for other Pc derivatives~\cite{kera_prb04,wang_acie09} -- could be confirmed
also by metastable atom electron spectroscopy (MAES)
measurements~\cite{hosokai_11}. Within our coverage series, we verified that a
deposition of more than $4\,$ML results in atomically disordered systems with
incoherent XSW signals ($f_H=0$) for all elements.

\begin{figure}[!t]
  \centering
  \includegraphics[width=0.7\columnwidth]{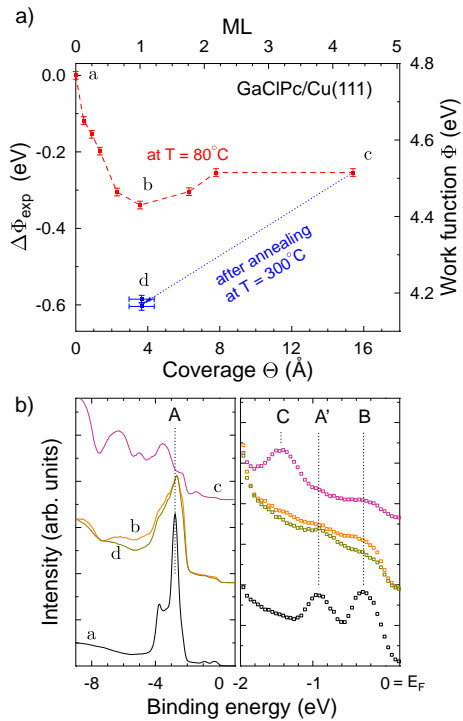}
  \caption{(a) Coverage dependent work function of  GaClPc on Cu(111). (b)
    Valence band region measured by UPS for coverages labeled correspondingly
    in the top panel.  In these spectra peak A belongs to the Cu\,3d band, A'
    is a satellite of A due to excitation by parasitic He\,I$_\beta$
    radiation, peak B is the Shockley surface state of the clean Cu(111), and
    peak C is the HOMO-derived state of the multilayer film.}
  \label{fig:workfunction_coverage_and_UPS}
\end{figure}

In order to study how the orientational order of the molecules affects the VL
shift we measured the work function of the system for increasing GaClPc
coverage (Fig.~\ref{fig:workfunction_coverage_and_UPS}a). At first, the
results obtained from the photoemission spectra reveal a rapid decrease below
$0.2\,$ML as expected for the 'Cl-down' configuration.  The continuous, yet
slightly retarded VL shift results in a minimum of $\Delta \Phi=-0.34\,$eV at
the monolayer coverage.
To overcome possible kinetic limitations we deposited more material (resulting
in a gradual increase of $\Phi$ for coverages of more than $1\,$ML due to
preferential 'Cl-up' nucleation) and applied a short annealing of the
multilayer at $300\,^\circ$C. This procedure yields a coverage of slightly
less than $1\,$ML and a shift of $\Delta \Phi = -0.60\,$eV. Because DFT
calculations, which we have performed using the adsorption geometry shown in
Fig.~\ref{fig:cu111_gaclpc1}c, predict a similar decrease of $\Delta \Phi =
-0.55\,$eV for a comparable coverage, we may relate the electronic properties
of annealed monolayer to the low-coverage XSW data.  Due to the orientational
order we can separate the total change $\Delta \Phi$ into two contributions
with
\begin{equation}
  \Delta \Phi = \Delta \Phi_\mathrm{dip} + \Delta \Phi_\mathrm{bond}.
\end{equation}
Thus, we distinguish $\Delta \Phi_\mathrm{dip}$, which is related to the
dipole moment of the molecules according to Eq.~(\ref{eq:dipole}), and $\Delta
\Phi_\mathrm{bond}$, which contains the effect of the molecule-metal
interaction.  To quantify the dipole contribution we have calculated a
monolayer of GaClPc without metal substrate. Again using the adsorption
geometry shown in Fig.~\ref{fig:cu111_gaclpc1}c we obtain $\Delta
\Phi_\mathrm{dip} = -0.30\,$eV.  Because of the large substrate-Pc distance
and the correspondingly small Pauli repulsion the associated value $\Delta
\Phi_\mathrm{bond} = -0.25\,$eV essentially reveals the effect of the charge
rearrangement within the Ga-Cl group and between the Cl and Cu
atoms~\cite{supp_mat}.

These conclusion are corroborated by the interpretation of the valence band
spectra shown in Fig.~\ref{fig:workfunction_coverage_and_UPS}b.  The data
obtained for different coverages show that neither before nor after the
annealing procedure prominent interface states -- as they were reported for
the adsorption of other planar~\cite{andreasson_sm08} and
non-planar~\cite{haemig_jesrp09} Pc molecules directly below the Fermi energy
-- can be observed. This absence of characteristic spectral features in the
HOMO-LUMO gap~\footnote{HOMO (LUMO): highest occupied (lowest unoccupied)
  molecular orbital} illustrates the weak interaction between the substrate
and the Pc ring in the 'Cl-down' orientation.
We note that due to the photoemission selection rules and the orientation of
the molecules no HOMO-derived states can be seen in the normal emission
spectra of both monolayer films shown
Fig.~\ref{fig:workfunction_coverage_and_UPS}b -- in contrast to the disordered
multilayer film, which exhibits a pronounced HOMO-related peak centered at
$-1.42\,$eV.
Moreover, the spectra illustrate the non-ideal growth of GaClPc at 80$\,
^\circ$C and the significant re-ordering caused by the annealing procedure.
The strong quenching of the Shockley surface state, which we find for the
annealed film, clearly indicates a more uniform substrate coverage.

In summary, the XSW data together with the UPS measurements and the DFT
calculations give a conclusive picture of the molecules' orientation and
bonding.  
For sub-monolayer coverages at sufficiently high substrate temperature the
vast majority of GaClPc molecules adsorb in the 'Cl-down' configuration with a
Cl-Cu layer spacing of $1.88\,$\AA{} that indicates covalent bonding of the
chlorine atoms.  The interaction with the substrate gives rise to subtle
deviations from the gas phase structure: In agreement with the DFT
calculations~\cite{supp_mat}, the Ga-Cl bond length of $2.33\,$\AA{} measured
by XSW is slightly larger than in the gas phase, while the average position of
the carbon atoms in the experiment and the calculations indicates a certain
bending of the molecule~\cite{supp_mat}.  Moreover, the coherent fractions
observed in the experiment can be related to thermal librations of the
molecules.  Due to the orientational order the molecular dipoles contribute
significantly to the vacuum level shift and the resulting interface dipole.
Detailed calculations for the annealed monolayer support this conclusion
quantitatively with $\Delta \Phi_\mathrm{dip} = -0.30\,$eV.  The influence of
intermolecular interactions is demonstrated by the coexistence of 'Cl-down'
and 'Cl-up' configurations found for deposition at 80$\,^\circ$C.
Interestingly, a subsequent annealing allows a controlled re-ordering of the
molecular dipoles and a change of the corresponding electronic properties.
Furthermore, we believe that the adsorption behavior of GaClPc and its impact
on the electronic structure represent important characteristics of non-planar
organic molecules.


We thank N.\ Koch and J.\ Pflaum for stimulating interactions, T.\ Suzuki and
H.\ Machida for experimental support.  We gratefully acknowledge the ESRF for
providing excellent facilities, the DFG (T\"ubingen), the G-COE (Chiba), and
the FWF (Graz) for funding our work.


\end{document}